\documentclass[pra,twocolumn,final,superscriptaddress,showpacs,aps]{revtex4-1}
\usepackage{amsmath}
\usepackage{amssymb}
\usepackage{graphicx}
\usepackage{color}

\begin{document}

\title{Numerical and variational solutions of the dipolar Gross-Pitaevskii equation in reduced dimensions}
\author{P. Muruganandam}
\affiliation{Instituto de F\'{\i}sica Te\'orica, UNESP - Universidade Estadual Paulista, 01.140-070 S\~ao~Paulo, S\~ao Paulo, Brazil}
\affiliation{School of Physics, Bharathidasan University, Palkalaiperur Campus, Tiruchirappalli 620024, Tamilnadu, India}
\author{S. K. Adhikari}
\affiliation{Instituto de F\'{\i}sica Te\'orica, UNESP - Universidade Estadual Paulista, 01.140-070 S\~ao~Paulo, S\~ao Paulo, Brazil}

\begin{abstract}

We suggest a simple Gaussian Lagrangian variational scheme for the {\it reduced} time-dependent quasi-one-  and quasi-two-dimensional  Gross-Pitaevskii (GP) equations of a dipolar Bose-Einstein  condensate (BEC) in cigar and disk configurations, respectively. The variational approximation for stationary states and breathing oscillation dynamics in reduced dimensions agrees well with the numerical solution of the GP equation
even for moderately large short-range and dipolar nonlinearities. The Lagrangian variational scheme also 
provides much physical insight about soliton formation  in dipolar BEC.

\end{abstract}

\pacs{03.75.Hh,03.75.Kk}

\maketitle
%\sloppy

\section{Introduction} 

The time-dependent mean-field Gross-Pitaevskii (GP) equation can accurately describe 
many static and dynamic  properties of a harmonically trapped 
Bose-Einstein condensate (BEC)  \cite{rev,rev-2,lpl-1,lpl-2,lpl-3,lpl-4,ref-a01,ref-a02,ref-a03,ref-a04,ref-b01,ref-b02,ref-b03,ref-b04,ref-b05,ref-b06,balaz2011}.
%, although 
%such a mean-field equation  is not expected  to reproduce properly
%true quantum effects like entanglement \cite{entanglement}, or scattering 
%dynamics \cite{scatt}. 
%Nevertheless, the GP equation has been consistently used in most  BEC phenomenology \cite{rev} where 
%the effect of quantum fluctuation is not important.  
However, the numerical solution of the three-dimensional (3D) GP 
equation could often be a   difficult task due to a large nonlinear term \cite{CPC,CPC-1}.  
Fortunately, in many experimental 
situations the 3D axially symmetric harmonic trap has extreme symmetry so that 
the BEC has either a cigar or a disk shape \cite{geom}. In these cases the essential 
statics and dynamics of a BEC take place in reduced dimensions. By integrating 
out the unimportant dimensional variable(s), reduced GP equations have been derived 
in lower dimensions \cite{luca,other,other-1}, which give a faithful description of the 
BEC in disk and cigar shapes.
For disk and cigar shapes the reduced GP equation is written in two (2D) and one dimensions
(1D), respectively. The numerical solution of such 2D, or 1D equation,  although  
simpler than that of the original 3D equation, remain  complex due to the nonlinear nature
of the GP equation.
Hence, for small values of the nonlinearity 
parameter, a Gaussian variational approximation is much useful for the solution of 
these equations \cite{var}.

The alkali metal atoms used in early BEC experiments have
negligible dipole moment. However, most bosonic atoms
and molecules have large dipole moments and a $^{52}$Cr
 \cite{lahaye,pfau,pfau-1,pfau-4}, and $^{164}$Dy \cite{dy,dy-2} BEC with a larger long-range dipolar interaction superposed
on the short-range atomic interaction, has been
realized. Other atoms, like $^{166}$Er \cite{otherdi,otherdi-1}, and molecules, such as  $^7$Li-$^{133}$Cs \cite{becmol},  with much larger dipole moment 
are being considered for BEC experiments.
A 3D GP equation for a dipolar BEC with a nonlocal nonlinear 
interaction  has been suggested \cite{lahaye} and successfully 
used to describe many properties of these condensates \cite{dipsol,jb,YY,Yi2000,Dutta2007-1,Dutta2007-2,Dutta2007-3,Dutta2007-5,Dutta2007-6}. 

The applicability of the nonlocal GP equation to the case of {dipolar BEC} has been a subject 
of intensive study \cite{lahaye}. After a detailed analysis, 
You and Yi \cite{YY,Yi2000} concluded that the  GP equation is valid for the {dipolar BEC}.
Further support on the validity of this equation came from the study of 
Bortolotti {\it et al.}
\cite{BB,BB-1}. 
 They compared the solution of the dipolar GP equation with the results of 
diffusive Monte Carlo calculations and found good agreement between the two.
However,  the 3D GP equation for a dipolar BEC 
with the nonlocal dipolar interaction has a complex structure and its 
numerical solution, involving the Fourier transformation of the dipolar 
nonlinear term to momentum space \cite{jb,YY,Yi2000}, 
is even more challenging than that of the GP equation 
of a non-{dipolar BEC}.  
  
Here we reconsider the dimensional reduction \cite{SS,deu}
of the 
GP equation to 1D form for  cigar-shaped dipolar BEC and obtain the 
precise  1D potential with a dipolar contact-interaction 
term.
%the 1D form of the reduced GP equation \cite{SS,deu}
%for cigar-shaped {dipolar BEC} with the    precise  1D potential with a 
%dipolar contact-interaction term. 
Previous derivations \cite{SS,deu}
of the 1D reduced equation for {dipolar BEC}
did not include the 
proper contact-interaction term, lacking which the 1D model will not
provide a correct description of the full 3D system.  
 We also   
consider  the reduced 2D GP
equation \cite{fisch,PS}  for a disk-shaped dipolar BEC.
%the dimensional reduction \cite{fisch,PS} of the 
%GP equation to 2D form for  disk-shaped {dipolar BEC}. 
Though these reduced GP equations for dipolar
BEC are computationally less expensive than their 3D counterparts, the numerical solution procedure
remains complicated due to 
repeated forward and backward Fourier transformations 
of the non-local dipolar term. As an alternative, 
here we
suggest  time-dependent 
Gaussian Lagrangian variational approximation of the 1D and 2D  reduced equations.
A direct attempt to derive the variational Lagrangian density of the reduced equations 
is not straightforward due to nonlocal integrals with error functions.
We present an indirect evaluation of the     Lagrangian density 
avoiding the above complex procedure.  Thus,
the present variational approximation  involves algebraical quantities 
without requiring any Fourier transformation to momentum space. 

In case 
of {dipolar BEC} of Cr and Dy atoms we consider the numerical solution of the 3D 
and the reduced 1D and 2D GP equations for cigar and disk shapes to 
demonstrate the appropriateness of the solution of the reduced 
equations. The  variational approximation of the reduced equations 
provided results for density, root-mean-square (rms) size, chemical 
potential, and breathing oscillation dynamics in good agreement with the 
numerical solution of the  reduced and full 3D GP equations.
 
\section{Analytical formulation} 

\subsection{3D GP Equation}

We  study a {dipolar BEC} 
of $N$ atoms, each of mass $m$, using the dimensionless 
GP
equation  \cite{pfau,pfau-1,pfau-4}
\begin{align}  \label{gp3d} 
i  \frac{\partial \phi({\bf r},t)}{\partial t}
 &\, =  \biggr[ -\frac{1}{2}\nabla^2 +V({\bf r}) + 4\pi a N\vert \phi({\bf r},t)\vert^2 \notag \\
&\, +  N \int U_{dd}({\bf r -r'})\vert\phi({\bf r'},t)\vert^2d^3{ r'}
\biggr] \phi({\bf r},t), \end{align} 
with  dipolar interaction 
$
 U_{dd}({\bf R}) = 3
a_{dd}(1-3\cos^2\theta)$ $/R^3,$ $\quad {\bf R=r-r'}.
$
 Here $V({\bf r})$ is the confining axially symmetric 
harmonic potential,
$\phi({\bf r},t)$  the 
wave function at time $t$ with normalization $\int \vert\phi({\bf r},t)\vert^2 d {\bf r}=1$, 
 $a$ the atomic scattering length, $\theta$ 
the angle between $\bf R$ and the  polarization direction  $z$.  
The constant $a_{dd}
=\mu_0\bar \mu^2 m /(12\pi \hbar^2)$ 
is a length characterizing the strength of 
dipolar interaction and its experimental
value for $^{52}$Cr  is $15a_0$ \cite{pfau,pfau-1,pfau-4}, with  $a_0$ the Bohr 
radius, 
 $\bar \mu$ the (magnetic) dipole moment of a single atom, and $\mu_0$ 
the permeability of free space.   
In equation  (\ref{gp3d}) length is measured in units of  characteristic 
harmonic oscillator length $l\equiv \sqrt{\hbar/m\omega}$,
angular frequency of trap in units of $\omega$,  
 time $t$
in units of $\omega^{-1}$, and energy in units of $\hbar\omega$.
The axial and radial angular frequencies of the trap are $\Omega_z\omega$
and $\Omega_\rho\omega$, respectively.   
The dimensionless 3D 
harmonic trap is  
\begin{equation}
 V({\bf r}) =  \frac{1}{2}\Omega_\rho^2\rho^2 + \frac{1}{2} \Omega_z^2 z^2,
\end{equation} 
where ${\bf r}\equiv (\vec\rho,z)$, with $\vec\rho$ the radial coordinate and $z$
the axial coordinate.
 
The Lagrangian density of equation 
  (\ref{gp3d})  is given by
\begin{align}
 {\mathcal L} &\,  =  \frac{i}{2}( \phi \phi^{\star}_t
- \phi^{\star}\phi_t)+\frac{\vert\nabla\phi\vert^2}{2}
+ V({\bf r})\vert\phi\vert^2 \notag \\ 
&\, + 2\pi aN\vert\phi\vert^4 
+  \frac{N}{2}\vert
\phi\vert^2\int U_{dd}({\mathbf r}-
{\mathbf r'})\vert\phi({\mathbf r'})\vert^2 d^3{  r}'
.\label{eqn:vari}
\end{align}
We use the Gaussian ansatz \cite{var,jb,YY,Yi2000} 
\begin{equation}\label{anz3d}
\phi({\bf r},t)= \frac{\pi^{-3/4}}{w_\rho \sqrt {w_z}} 
\exp\left(-\frac{\rho^2}{2w_\rho^2} - \frac{z^2}{2w_z^2} +i\alpha\rho^2
+i\beta z^2 \right) 
\end{equation}
for a variational calculation, 
where $w_\rho$ and $w_z$ are time-dependent radial and axial widths,
and $\alpha$ and $\beta$  time-dependent phases. 
The effective Lagrangian $L\equiv 
 \int {\mathcal L}d^3{  r}$ (per particle) becomes
\begin{eqnarray}
L & = &\,
  \left(w_\rho^2\dot{\alpha} +
\frac{w_z^2\dot{\beta}}{2}\right) %-2V_\rho \exp(-w_\rho^2)
+\frac{\Omega_\rho^2 w_\rho^2}{2}
+\frac{\Omega_z^2 w_z^2}{4}
+ \frac{1}{2{w_\rho^2}} + \frac{1}{4w_z^2}
\nonumber \\ &&\,
+ 2w_\rho^2 \alpha^2 + w_z^2\beta^2 
+ \frac{N  }{(\sqrt{2 \pi}
w_\rho^2w_z)} \left[ {a}-{a_{dd}} 
f(\kappa)\right], \label{lag:eff}
\end{eqnarray}
with 
\begin{align} & f(\kappa)= \frac{1+2\kappa^2-3\kappa^2d(\kappa)}{(1-\kappa^2)},  \\
& d(\kappa)= \frac{\mbox{atanh}\sqrt{1-\kappa^2}}{\sqrt{1-\kappa^2}}, \;\;
\kappa=\frac{w_\rho}{w_z}.
\end{align}
The Euler-Lagrange equations for variational parameters $w_\rho, w_z, \alpha$ and $\beta$ yield
the following equations for widths
 $w_\rho$ and $w_z$ 
 \begin{eqnarray} &&
\ddot{w}_{\rho}%+\frac{4V_\rho  {w_\rho}}{e^{w_\rho^2}}
+\Omega_\rho^2 w_\rho 
=
\frac{1}{w_\rho^3} +\frac{
N}{\sqrt{2\pi}} \frac{  \left[2{a} - a_{dd}
{g(\kappa) }\right]  }{w_\rho^3w_{z}}
,
\label{f1} \\ && \ddot{w}_{z} 
 +\Omega_z^2 w_z =
\frac{1}{w_z^3}+ \frac{ 2N}{\sqrt{2\pi}}
\frac{ \left[{a}-a_{dd}
c(\kappa)\right]  }{w_\rho^2w_z^2} , \label{f2} 
\end{eqnarray}
with 
\begin{align}
& g(\kappa)=\frac{2-7\kappa^2-4\kappa^4+9\kappa^4 d(\kappa)}{(1-\kappa^2)^2}, \\ 
& c(\kappa) =\frac{1+10\kappa^2 -2\kappa^4 -9\kappa^2 d(\kappa)}{(1-\kappa^2)^2}.
\end{align}
The chemical potential $\mu$ for a stationary state is 
\begin{align}
\mu =  &\, \frac{1}{2w_\rho^2}+
\frac{1}{4w_z^2}+\frac{2N[a-a_{dd}f(\kappa)]}{\sqrt{2\pi} w_zw_\rho^2} 
%\notag \\ &\,
+\frac{\Omega_\rho^2 w_\rho^2}{2}
+\frac{\Omega_z^2 w_z^2}{4}.
\end{align}

\subsection{1D reduction} 

For a cigar-shaped {dipolar BEC} with a strong radial trap $(\Omega_\rho
>\Omega_z) $  
one can write the following effective
%We can now derive the effective
%1D equation for the cigar-shaped {dipolar BEC}. We substitute the
%ansatz (\ref{anz1d}) in equation   (\ref{gp3d}), multiply by the ground-state wave
%function $\phi(\rho)$ and integrate in $\rho$ to get
 1D equation (details given in Appendix)
\begin{align}\label{gp1d}
& i\frac{\partial \phi_{1D}(z,t)}{\partial
t}=\biggr[-\frac{\partial_z^2}{2}+\frac{\Omega_z^2 z^2}{2}+ 
\frac{2 aN}
{
d_\rho^2}\vert\phi_{1D}\vert^2+ \frac{2a_{dd}N}{d_\rho^2}
\nonumber \\ &\times
\int_{-\infty}^{\infty}\frac{dk_z}{2\pi}e^{ik_z z}\tilde
n(k_z)s_{1D}\left(\frac{k_z d_\rho}{\sqrt 2}\right)\biggr]\phi_{1D}(z,t) ,
%\\
%&\equiv  \biggr[-\frac{\partial_z^2}{2}+\frac{\Omega_z^2 z^2}{2}+
% \frac{2 a N\vert\phi_{1D}\vert^2}
%{
%d_\rho^2} \nonumber \\
%& + N\int_{-\infty}^{\infty} U_{dd}^{1D}( Z )\vert\phi_{1D}({z'},t)\vert^2d{z'}
%\biggr] \phi_{1D}(z,t).
\end{align} 
where $s_{1D}$ is defined by equation  (\ref{zeta}) and  $d_\rho\equiv 1/\sqrt{\Omega_\rho}$ is the radial harmonic oscillator length.

 To solve equation  (\ref{gp1d}), we use 
 the Gaussian variational ansatz 
\begin{equation}\label{anzv1d}
\phi_{1D}(z)= \frac{\pi^{-1/4}}{\sqrt{w_z}}\exp\left[-\frac{z^2}{2w_z^2}+i\beta z^2\right]  .
\end{equation} 
From equation  (\ref{anz1d}) we see that the variational 1D ansatz (\ref{anzv1d})
corresponds to the 
following 3D wave function
\begin{equation}\label{1d3d}
\phi({\bf r},t)= \frac{\pi^{-3/4}}{d_\rho \sqrt {w_z}} 
\exp\left(-\frac{\rho^2}{2d_\rho^2} - \frac{z^2}{2w_z^2} +i\beta z^2
  \right) .
\end{equation} 
The present variational 
wave function (\ref{1d3d}) is a special case of the 3D variational wave function 
(\ref{anz3d}) with $w_\rho=d_\rho$ and $\alpha=0$.   Hence,
the 1D variational Lagrangian can be written from the 3D Lagrangian 
(\ref{lag:eff}), (using $w_\rho=d_\rho$ and $\alpha=0$,) as  
\begin{align}
 L_{1D}  & =  \frac{w_z^2\dot{\beta}}{2} 
 %-V_z\exp(-w_z^2)
 + \frac{1}{4w_z^2}
+ w_z^2\beta^2+\frac{\Omega_z^2 w_z^2}{4}  \notag\\
& + \frac{N  }{\sqrt{2 \pi} d_\rho^2w_z} \left[ {a}-{a_{dd}} 
f(\kappa_0)\right]
; \quad  \kappa_0=\frac{d_\rho}{w_z}, \label{1dlag}
\end{align}
where we have removed the constant terms.
{ This inductive derivation of the 1D Lagrangian (\ref{1dlag})
avoids the construction of Lagrangian density involving error functions
in the 1D potential (\ref{1dpotx}) and subsequent integration to 
obtain the Lagrangian.}
The Euler-Lagrange equation for the variational parameter 
$w_z$  of Lagrangian (\ref{1dlag}) is
\begin{eqnarray}\ddot w_z + 
\Omega_z ^2 w_z
%+\frac{4V_zw_z}{\exp(w_z^2)} 
= \frac{1}{w_z^{3}}
+\frac{2 N[a-a_{dd}c(\kappa_0)]}
{\sqrt
{2\pi}w_z^2 d_\rho^2}. \label{v1d}
\end{eqnarray}
The variational 
chemical potential  is given by 
\begin{eqnarray}\mu=
\frac{1}{4w_z^2}+\frac{2N[a-a_{dd}f(\kappa_0)]}{\sqrt{2\pi} w_z d_\rho^2}
+\frac{\Omega_z^2 w_z^2}{4 }.
% -\frac{V_z}{\exp(w_z^2)}. 
\end{eqnarray}

Not only are the above variational results simple and   yield a good approximation to  
the 1D GP equation, much physical insight 
about the system can be obtained from the  variational Lagrangian  (\ref{1dlag}). In 
a quasi-1D system, the axial width is much larger than the transverse oscillator 
length:
$w_z\gg d_\rho$. Consequently, $\kappa_0\to 0$ and $f(\kappa_0)\to 1$. From equation  (\ref{1dlag}), we 
see that the interaction term becomes in this limit $N(a-a_{dd})/(\sqrt{2 \pi} d_\rho^2w_z)$. In 
equation  (\ref{gp1d}), the dipolar term involves a nonlocal integral. However, the variational approximation
suggests that the effect of the dipolar interaction integral  is to reduce the contact interaction term 
in equation  (\ref{gp1d}) replacing the scattering length $a$ by $(a-a_{dd})$. Immediately, one can conclude 
that the system effectively becomes attractive for $a_{dd}>a$.  So one can have the formation of 
bright soliton even for positive (repulsive) scattering length $a$, provided that $a_{dd}>a$. 
%The variational sound velocity in this case is $c_{1D}=\sqrt{2n_{1D} (a-a_{dd})/d_\rho^2}$ \cite{rev}, 
%consistent with the result of Bogoliubov theory \cite{pre},
%where $n_{1D}$ is the 1D density. It shows that with the increase of $a_{dd}$, sound velocity is reduced 
%and becomes imaginary for $a_{dd}>a$, corresponding to no sound. These nontrivial 
%conclusions about the existence 
%of bright soliton \cite{dbs}
%and sound propagation \cite{pre} has been found after solving the dipolar GP equations in 
%1D and 3D.  

\subsection{2D reduction} In the  disk-shape,  with a 
strong axial trap ($\Omega_z>\Omega_\rho$),  
the {dipolar BEC} is assumed to be in the ground state $\phi(z)= \exp(-z^2/2d_z^2)/{(\pi d_z^2)}^{1/4}$
of the axial trap and  the wave function 
$\phi({\bf r}) $ %=\phi_{2D}(x,y)  \phi(z)$
can be written
as \cite{fisch,PS}
\begin{equation}\label{anz2d}
\phi({\bf r})= \frac{1}{{(\pi d_z^2)}^{1/4}}\exp \left(-\frac{z^2}{2d_z^2}\right) \phi_{2D}(x,y),
\end{equation}
where $ \phi_{2D}(x,y)$ is the 2D wave function and $d_z=\sqrt{1/\Omega_z}$. 
Using  ansatz (\ref{anz2d}) in equation   (\ref{gp3d}), the $z$ dependence can be integrated out 
to obtain the following effective  2D equation  \cite{fisch,PS}
\begin{align}
\label{gp2d}
&\, i\frac{\partial \phi_{2D}(\vec \rho,t)}{\partial t} = 
\biggr[-\frac{\nabla_\rho^2}{2}
+\frac{\Omega_\rho^2\rho^2}{2} +\frac{4\pi a N}{\sqrt{2\pi}d_z}
\vert\phi_{2D}\vert^2 +\frac{4\pi a_{dd} N}{\sqrt{2\pi}d_z} \nonumber \\
&\, \quad \times \int \frac{d^2 k_\rho}{(2\pi)^2}
\exp ({i{\bf k}_\rho \cdot {\vec \rho}}) {\tilde n}({\bf k_\rho})
h_{2D}(\frac{k_\rho d_z}
{\sqrt 2}) \biggr]  \phi_{2D}(\vec \rho,t), \\
&\, \tilde n ({\bf k}_\rho)  = \int \exp \left(i {\bf k}_\rho \cdot \vec \rho \right) 
|\phi_{2D}(\vec\rho) |^2 d \vec\rho,
\end{align}
where $h_{2D}(\xi) = 2-3\sqrt\pi \xi e^{\xi^2}{\mbox{erfc}}(\xi)$  \cite{PS}, 
${\bf k}_\rho \equiv (k_x, k_y)$, and the dipolar term is written   in  Fourier  space.

To solve  equation  (\ref{gp2d}), we use the Gaussian ansatz 
\begin{equation}\label{anzv2d}
\phi_{2D}(\rho)=\frac{1}{w_\rho\sqrt \pi}\exp\left(-\frac{\rho^2}{2w_\rho^2}+i\alpha \rho^2\right).
\end{equation} 
From equation  (\ref{anz2d}) we see that the 2D wave function (\ref{anzv2d})
corresponds to the 
following 3D wave function
\begin{equation}\label{2d3d}
\phi({\bf r},t)= \frac{\pi^{-3/4}}{w_\rho \sqrt {d_z}} 
\exp\left(-\frac{\rho^2}{2w_\rho^2} - \frac{z^2}{2d_z^2} +i\alpha\rho^2
  \right) .
\end{equation} 
The present variational 
wave function (\ref{2d3d}) is a special case of the 3D variational wave function 
(\ref{anz3d}) with $w_z=d_z$ and $\beta=0$.   Hence,
the 2D variational Lagrangian can be written from the 3D Lagrangian 
(\ref{lag:eff}) as 
\begin{align}\label{2dlag}
 L_{2D}  &\, =  {w_\rho^2\dot{\alpha}} +\frac{ w_\rho^2\Omega_\rho^2}{2}
+ \frac{1}{2w_\rho^2}
+ 2w_\rho^2\alpha^2 \notag \\
 &\,+ \frac{N  }{\sqrt{2 \pi} w_\rho^2d_z} \left[  {a}-{a_{dd}} 
f(\bar \kappa)\right]; \quad  \bar \kappa=\frac{w_\rho}{d_z},
\end{align}
where we have removed the constant terms.
The Euler-Lagrange 
variational equation for width $w_\rho$ becomes 
\begin{eqnarray}  
\ddot{w}_{\rho}+ 
{w_\rho\Omega_\rho^2}
=
\frac{1}{w_\rho^3} +\frac{
N}{\sqrt{2\pi}} \frac{  \left[2{a} - a_{dd}
{g(\bar \kappa) }\right]  }{w_\rho^3d_{z}}.
\label{v2d}  
\end{eqnarray}
The chemical potential $\mu$ for a stationary state is 
\begin{eqnarray}
\mu& = & \frac{1}{2w_\rho^2}+
\frac{2N[a-a_{dd}f(\bar \kappa)]}{\sqrt{2\pi} d_zw_\rho^2} 
+\frac{ w_\rho^2\Omega_\rho^2}{2}.
\end{eqnarray}

In a quasi-2D system, the radial width is much larger than the axial oscillator 
length:
$w_\rho\gg d_z$. Consequently, $\bar \kappa\to  \infty$ and $f(\bar \kappa)\to -2$. From equation (\ref{2dlag}), we 
see that the interaction term becomes in this limit $N(a+2a_{dd})/(\sqrt{2 \pi} w_\rho^2d_z)$.  
The variational approximation
suggests that the effect of the dipolar interaction in equation  (\ref{gp2d})  is to increase the contact interaction term 
 replacing  $a$ by $(a+2a_{dd})$. 
%Thus, 
% the system effectively becomes repulsive for $a_{dd}>|a/2|,$  $a<0$.  
Hence, for positive $a$, there cannot be any bright soliton in 2D, which 
was found from a solution of the 2D GP  equation (\ref{gp2d}) and Bogoliubov theory \cite{tick}.
However, effectively the sign of the 
dipolar term in the GP equation can be changed by  rotating the external field that orients the dipoles much
faster than any other relevant time scale in the system  \cite{nath}. In this fashion Nath {\it et al.} \cite{nath}
suggest changing the dipole interaction term by a factor of $-1/2$, which changes the effective scattering 
length in the Lagrange variational approximation to $(a-a_{dd})$, (as discussed in the 1D case above,) leading to the formation of bright 2D solitons for $a_{dd}>a$. These solitons were obtained  by Nath {\it et al.} from a  solution of the 2D GP equation (\ref{gp2d}).

\section{Numerical results} We solve the 1D, 2D, and 3D  GP equations 
employing imaginary- and real-time propagation with Crank-Nicolson 
method \cite{CPC,CPC-1}.  
The dipolar interaction is
evaluated by  fast Fourier transform \cite{jb,YY}.

We present results for $^{52}$Cr and $^{164}$Dy atoms. The  $^{52}$Cr has a 
moderate dipole moment with $a_{dd}=15a_0$ \cite{pfau,pfau-1,pfau-4}, 
while the   $^{164}$Dy atom has a large dipole 
moment with  $a_{dd}=130a_0$  \cite{dy,dy-2}.  In both cases we present results 
for {dipolar BEC} of up to 10,000 atoms
for $0 < a < 10$ nm and   
choose the  frequency $\omega$ such that the oscillator  length $l=1 \mu$m.

First, we present the results for density profiles obtained from a 
solution of the reduced 1D and 2D equation and compare with the full 3D 
results. It is known that the densities obtained from the reduced equations agree 
well with the full 3D density, as the nonlinearity tends to zero and/or 
the trap asymmetry is extreme \cite{luca}. Hence in this study we 
consider a moderately small trap asymmetry and a relatively large 
nonlinearity of experimental interest.  In the cigar (1D) case we 
consider $^{52}$Cr atoms with $a=6$ nm, and in the disk (2D) case we 
consider $^{164}$Dy atoms with $a=6$ nm.

\begin{figure}[!ht]
\begin{center}
\includegraphics[width=.99\linewidth]{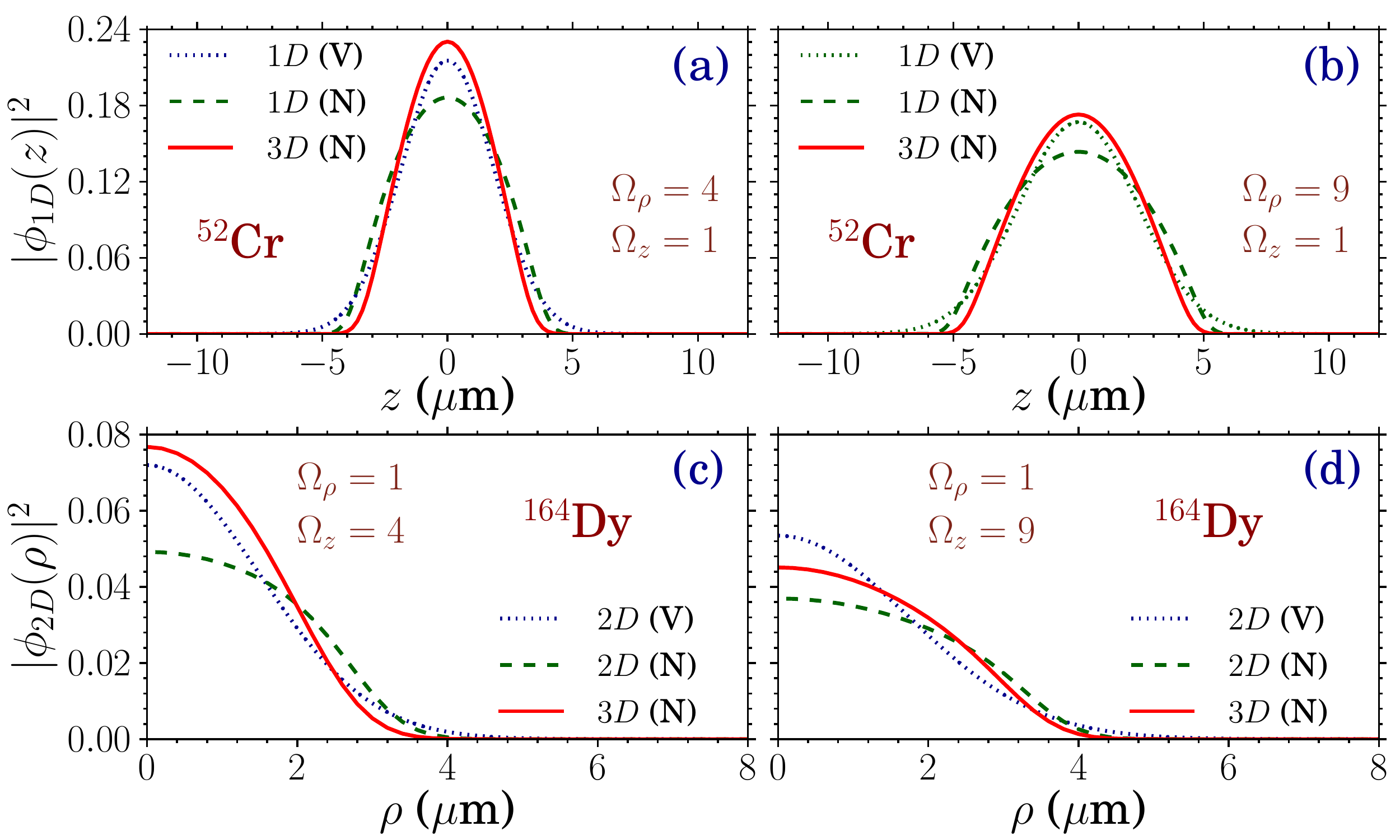}
\end{center}

\caption{ Linear density of a cigar-shaped $^{52}$Cr {dipolar BEC} of 1,000
atoms of $a=6$ nm, with trap parameters $\Omega_z=1$ and (a)  $\Omega_\rho =4, $
and (b)  $\Omega_\rho =9 $ from a numerical (N) solution of the 3D equation 
(\ref{gp3d}) and 1D equation (\ref{gp1d}), and its variational (V) result. 
Radial  density of a disk-shaped $^{164}$Dy {dipolar BEC} of 1,000
atoms of $a=6$ nm, with trap parameters $\Omega_\rho=1$ and (c)  $\Omega_z =4, $
and (d)  $\Omega_z =9 $
  from a numerical  solution of the 3D equation 
(\ref{gp3d}) and  2D equation (\ref{gp2d}), and its variational result. 
}

\label{fig1}
\end{figure}

In Figs.   \ref{fig1} (a) and (b), we plot results for linear density of 
a cigar-shaped $^{52}$Cr {dipolar BEC} of 1,000 atoms 
as calculated from the numerical 
solution of the 3D equation (\ref{gp3d}) and the 1D equation 
(\ref{gp1d}) and its variational result (\ref{v1d}) for $\Omega_z=1$ 
and $\Omega_\rho = 4$ and 9. We find,  as the trap asymmetry increases by 
changing $\Omega_\rho$ from 4 to 9, the agreement between 3D and 1D 
models improves. In Figs.  \ref{fig1} (c) and (d), we plot results for 
radial density of a disk-shaped $^{164}$Dy {dipolar BEC} of 1,000 atoms
as calculated from 
the numerical solution of the 3D equation (\ref{gp3d}) and the 2D 
equation (\ref{gp2d}) and its variational approximation (\ref{v2d}) for 
$\Omega_\rho=1$ and $\Omega_z = 4$ and 9. We find that, with the 
increase of the trap asymmetry from $\Omega_z =4$ to 9, the agreement 
between the 3D and 2D models enhances. In all cases the variational 
results of the reduced 1D and 2D equations are in good agreement with 
those of the full 3D model.

After having established the appropriateness of the reduced 1D and 2D 
equations in the cigar and disk shapes, it is realized that although the 
numerical solution of these reduced GP equations are simpler than that 
of the full 3D GP equation, they are still complicated due to the 
presence of the nonlocal dipolar interaction. The variational approximation
of these equations presented here is relatively
simple and could be used for approximate solution of these equations. 
Now we test the variational results of the reduced 1D and 2D equations 
by comparing with the numerical%The contribution of the dipole potential to energy is
%\begin{eqnarray}
%\label{poten}
%H_{dd} & = &\frac{N}{2} \int d^3 r\int d^3 r'n({\bf r}) U_{dd}({\bf
%r-r'}) n({\bf r'}), \\
%&=& \frac{1}{2} \frac{N}{(2\pi)^3}\int d^3 k \tilde n ({\bf k})
%\tilde U_{dd} ({\bf k})\tilde n (-{\bf k}),\label{ftpot}
%\end{eqnarray}
%where $n({\bf r})\equiv |\phi({\bf r})|^2 $ is the density and 
%in equation  (\ref{ftpot})
%we used a convolution of the respective variables 
%to Fourier space and where tilde denotes  
%Fourier
%transformations: \cite{jb,fisch,PS}
%\begin{eqnarray}
%\tilde U_{dd} ({\bf k})&=&\frac{4\pi}{3}3  a_{dd} \biggr[ 
%\frac{3k_z^2}{k^2}-1 \biggr], \\
% \tilde n ({\bf k})&=& \exp (-k_z^2 d_z^2/4) \tilde n_{2D} (k_x,k_y).\label{ftwf}
%\end{eqnarray}
 solution of these equations.

\begin{figure}[!ht]
\begin{center}
\includegraphics[width=.99\linewidth]{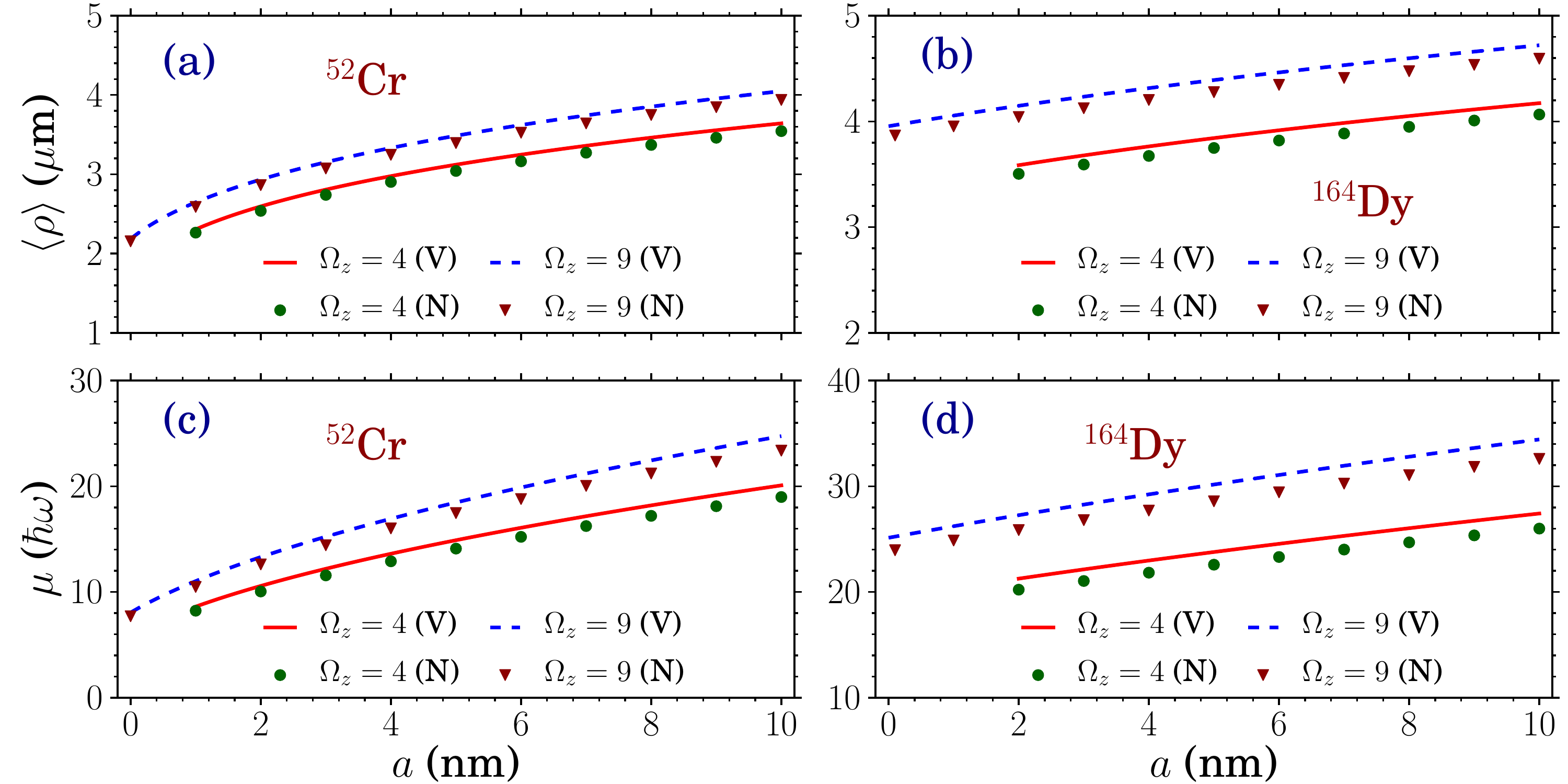}
\end{center}

\caption{The numerical (N) and variational (V) rms size $\langle \rho \rangle $  
versus scattering length $a$ of a disk-shaped {dipolar BEC} of 10,000  (a) $^{52}$Cr and 
(b) $^{164}$Dy atoms for trap parameters $\Omega_\rho =1$ and $\Omega_z=4$ and 9
from a solution of the reduced 2D GP equation (\ref{gp2d}).
The corresponding chemical potential $\mu$ in these cases for  (c) $^{52}$Cr and 
(d) $^{164}$Dy atoms.
}

\label{fig2}
\end{figure}

\begin{figure}[!ht]
\begin{center}
\includegraphics[width=.99\linewidth]{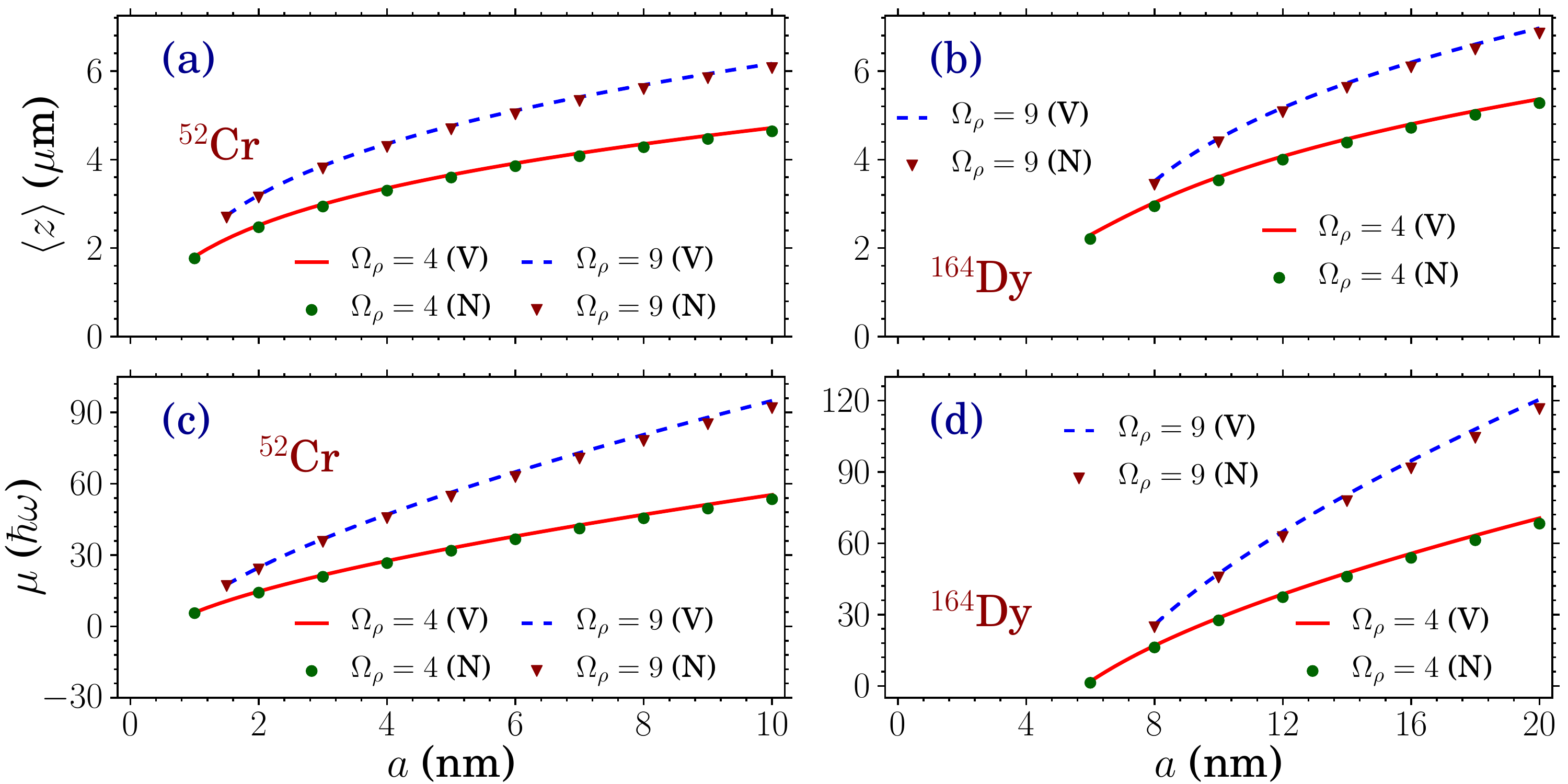}
\end{center}

\caption{The numerical (N) and variational (V) rms length  $\langle z \rangle $  
versus scattering length $a$ of a cigar-shaped {dipolar BEC} of 10,000  (a) $^{52}$Cr and 
(b) $^{164}$Dy atoms for trap parameters $\Omega_z =1$ and $\Omega_\rho=4$ and 9
from a solution of the reduced 1D GP equation (\ref{gp1d}).
The corresponding chemical potential $\mu$ in these cases for  (c) $^{52}$Cr and 
(d) $^{164}$Dy atoms.
}

\label{fig3}
\end{figure}

In Figs.  \ref{fig2} we present the results for rms size $\langle \rho 
\rangle$ and chemical potential $\mu$ of a disk-shaped $^{52}$Cr and 
$^{164}$Dy {dipolar BEC} of 10,000 atoms with the trap parameters $\Omega_\rho=1$ 
and $\Omega_z=4$ and 9 for $0 < a < 10$ nm as calculated from numerical 
and variational approaches of the reduced 2D equation (\ref{gp2d}).  In 
Figs.    \ref{fig3} we exhibit the results for rms size $\langle z 
\rangle$ and chemical potential $\mu$ of a cigar-shaped $^{52}$Cr and 
$^{164}$Dy {dipolar BEC} of 10,000 atoms with the trap parameters $\Omega_z=1$ 
and $\Omega_\rho=4$ and 9 for $0 < a < 20$ nm as calculated from 
numerical and variational approaches of the reduced 1D equation 
(\ref{gp1d}).

The dipolar interaction changes from strongly attractive in the extreme 
cigar shape ($\Omega _\rho \gg \Omega_z$) to strongly repulsive in the 
extreme disk shape ($\Omega _\rho \ll \Omega_z$) and its effect is 
minimum (nearly zero) for $\Omega _\rho$ slightly less than $\Omega _z$. 
In Fig. \ref{fig2} the dipolar interaction is slightly attractive for 
$\Omega_z=4$ and $\Omega_\rho=1$. Hence in the absence of any 
short-range interaction ($a=0$), the system will collapse and no stable 
solution of the GP equation can be obtained. For Cr atoms the dipolar 
interaction is weak, and for $a\ge 1$ nm, the short-range repulsion for 
10,000 atoms {surpluses} the dipolar attraction and a stable state can 
be obtained for $\Omega_z=4$.  For Dy atoms the dipolar interaction is 
stronger, and a stable state can be obtained only for $a\ge 2$ nm for 
$\Omega_z=4$. For $\Omega_z=9$, the dipolar interaction for both Cr and 
Dy atoms are repulsive and a stable state is obtained in this case for 
$a>0$. In Fig.  \ref{fig3} the dipolar interaction is attractive for 
both $\Omega_\rho=4$ and 9. Hence the {dipolar BEC} can be stable only for 
scattering length $a$ greater than a critical value. This is why the 
curves in this figure start above this critical value. This critical 
value is larger for Dy atoms and $\Omega_\rho=9$ compared to that of Cr 
atoms and $\Omega_\rho=4$ as can be found in Fig. \ref{fig3}. As there 
is no real collapse in 1D models with cubic nonlinearity; for confirming 
the collapse correctly one must solve the full 3D GP equation.

%{
% The relative differences in percentage
%between the numerical and
%variational rms sizes and chemical potentials
%illustrated  in Figures   \ref{fig2} and \ref{fig3} are shown in figure \ref{fig4} (a),
%(b), (c), and (d), where we present the percentage difference for $\langle \rho \rangle$ and $\mu$ 
%for the 2D model of  figure \ref{fig2} and for $\langle z \rangle$ and $\mu$ for 
%the 1D model of  figure \ref{fig1}.
%This shows that the variational results lie within few percent of the numerical 
%results for rms size and chemical potential. 
%In all cases the agreement between the numerical and 
%variational results is good, which confirms the adequateness of the 
%variational approximation. 
%}
%\begin{figure}[!ht]
%\begin{center}
%\includegraphics[width=.99\linewidth]{fig4.pdf}
%\end{center}
%
%
%\caption{  Relative differences (\%) between the numerical and variational results  
%of (a) the  rms  size $\langle \rho \rangle$,  (b) the chemical potential 
%$\mu$ shown in figure \ref{fig2}. The same for (c) $\langle z \rangle$ and (d) $\mu$ 
%shown in figure \ref{fig3}.   }
%
%
%\label{fig4}
%\end{figure}

\begin{figure}[!ht]
\begin{center}
\includegraphics[width=.99\linewidth]{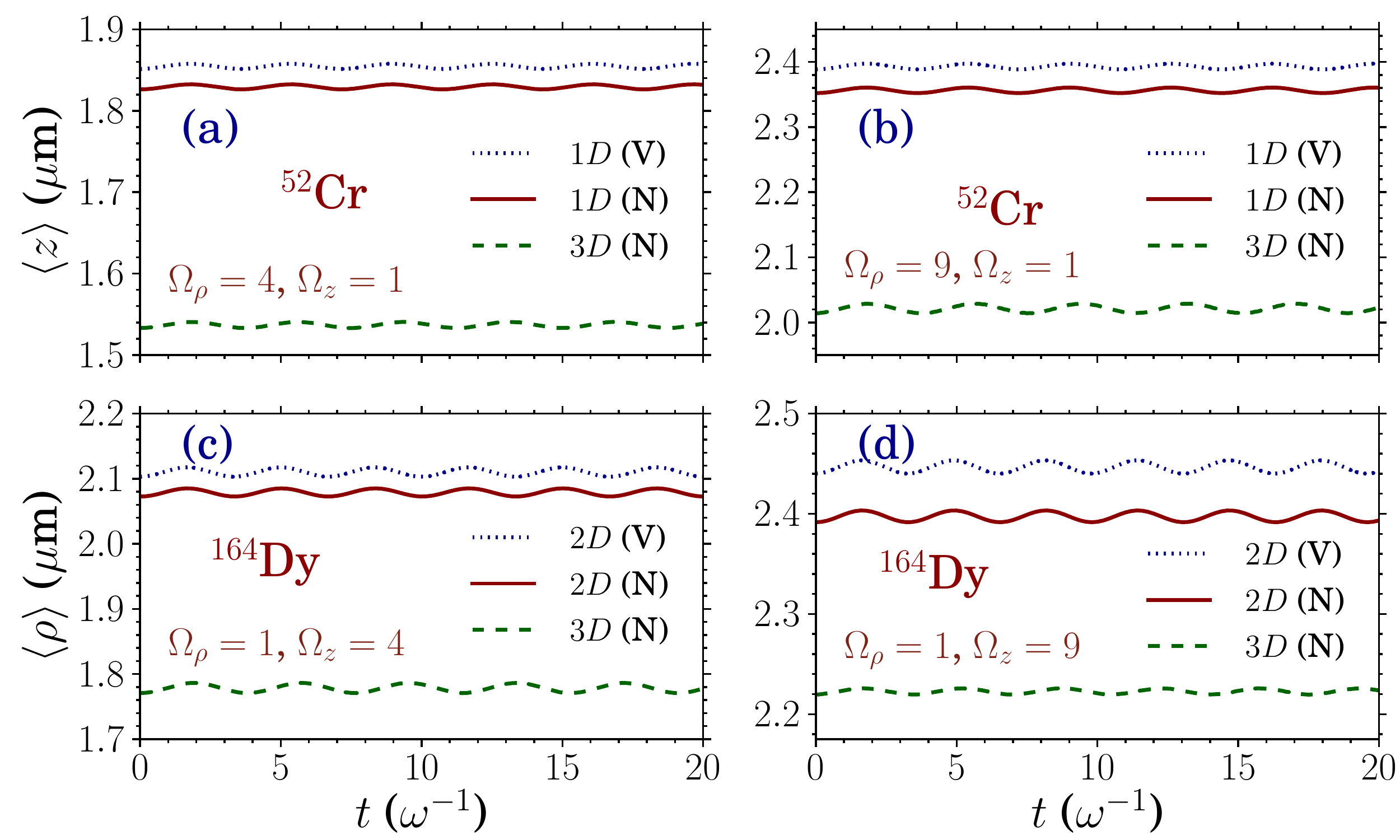}
\end{center}

\caption{ The rms sizes $\langle z \rangle $  of a  cigar-shaped Cr {dipolar BEC} 
of 1,000 atoms versus time $t$  for (a) $\Omega_\rho =4$ and (b) 9
from a numerical (N) and variational (V) results of the reduced 1D equation. 
The rms sizes   $\langle \rho \rangle $  of a  disk-shaped Dy {dipolar BEC} 
of 1,000 atoms versus time $t$ for (c) $\Omega_z =4$ and (d) 9. 
The oscillation was initiated by jumping the scattering 
length $a$ from 6 nm to  $6.15$ nm for Cr ($6.3$ nm for Dy) at $t=0$ 
from a solution of the reduced 2D 
equation.
 }

\label{fig5}
\end{figure}

Next we study, by numerical and variational solutions of the reduced 1D 
and 2D equations, the dynamics of breathing oscillation of the four {dipolar BEC} 
of cigar- and disk-shaped Cr and Dy atoms shown in Figs.  \ref{fig1} 
started by a small change of the scattering length. This can be 
implemented experimentally by a Feshbach resonance \cite{fesh}. In 
Fig.  \ref{fig5} this dynamics is shown for a cigar-shaped Cr {dipolar BEC} of 
1,000 atoms for (a) $\Omega_\rho =4$ and (b) $\Omega_\rho =9$ from a 
solution of the reduced 1D equation,
 and for a disk-shaped Dy {dipolar BEC} of 1,000 atoms for (c) $\Omega_z =4$ and 
(d) $\Omega_z =9$ from a solution of the reduced 2D equation. In these 
figures   we also show the results from a numerical solution of the 3D 
Eq. (\ref{gp3d}). The agreement between the numerical and 
variational results is good in all cases. We also calculated the angular 
frequencies of these oscillations. In case of Cr in Figs.    \ref{fig5} 
(a) and (b), the axial frequencies are 1.75 (variational, 1D), 1.76 
(numerical, 1D) and 1.63 (numerical, 3D), and in case of Dy in  Figs.   
\ref{fig5} (c) and (d), the radial frequencies are 1.93 (variational, 
2D), 1.89 (numerical, 2D) and 1.76 (numerical, 3D). For quasi-linear 
systems, these angular frequencies are expected to be 2 
\cite{stringari}. The deviation from this value is due to the large 
nonlinearity of the {dipolar BEC}s considered here.

\section{Conclusion} The usual GP equation provides a good description of statics and dynamics of
a normal  non{dipolar BEC}. For a {dipolar BEC} the numerical solution of the GP equation is a  
difficult task 
due to the nonlocal dipolar interaction.   For a cigar- and disk-shaped {dipolar BEC}, the reduced 1D and 2D 
equations provide an alternative to the full 3D equation. Nevertheless, the solution of these reduced 
equations is also challenging involving Fourier and inverse Fourier transformations. As an alternative, 
we suggest a time-dependent 
variational scheme for   these reduced equations, not requiring any Fourier transformation.
The variational approximation of these reduced equations provides results for stationary  cigar- and disk-shaped {dipolar BEC}
as well as for breathing oscillation of the same in good agreement with the numerical solution of the 
respective GP equations. This is illustrated for  large Cr and Dy  {dipolar BEC}s of 10,000 atoms and large atomic 
scattering lengths $a$ up to 20 nm. We also study the breathing oscillation of a bright soliton of 1,000 Cr 
atoms using the numerical solution of the 3D equation as well as the numerical and variational approaches 
to the 1D equation. 
A typical {dipolar BEC} considered here  corresponds to a large short-range cubic nonlinearity of about 
$4\pi a N \approx 1250$ for $a=10$ nm and $N=10,000$ and 
a large dipolar nonlinearity of $4\pi a_{dd}\approx  865$ for Dy atoms
for $a_{dd}=130a_0$ and $N=10,000$. The variational approximations considered here 
provided good results for such large nonlinearities and should be useful for analyzing the statics and 
dynamics 
of realistic {dipolar BEC}s 
under appropriate experimental  conditions.

%\acknowledgement
%\section*{Acknowledgments}
%\section*{Acknowledgments}
\acknowledgements
We thank FAPESP (Brazil), CNPq (Brazil), DST (India), and CSIR (India) for partial support.

%\section{1D reduction}
\appendix
\section{1D reduction}

For  a cigar-shaped {dipolar BEC} with a
strong radial trap ($\Omega_\rho > \Omega_z$), we assume that in the
radial direction the {dipolar BEC} is confined in the ground state
\begin{align}
\phi({\bf \rho}) = \exp(-(\rho^2/2d_\rho^2)/(d_\rho\sqrt \pi)
\end{align}
 of the transverse trap and the wave function   $\phi({\bf r})
 = \phi_{1D} (z) $ $\times \phi({\bf \rho})$ can be written as \cite{SS,deu}
\begin{align}\label{anz1d}
\phi({\bf r}) =  \frac{1}{\sqrt{\pi d_\rho^2}}\exp \left[-\frac{\rho ^2}
{2d_\rho^2}\right] \phi_{1D}(z);\quad  \Omega_{\rho}d_\rho^2=1,
%\Omega_{\rho}^2d_\rho^4&=&1+(g-{\frac{4\pi}{3}g_d})n/(2\pi),
\end{align}
where $d_\rho$ is the radial harmonic oscillator length.

The contribution of the dipole potential to energy is
\begin{eqnarray}
\label{poten}
H_{dd} & = &\frac{N}{2} \int d^3 r\int d^3 r'n({\bf r}) U_{dd}({\bf
r-r'}) n({\bf r'}), \\
&=& \frac{1}{2} \frac{N}{(2\pi)^3}\int d^3 k \tilde n ({\bf k})
\tilde U_{dd} ({\bf k})\tilde n (-{\bf k}),\label{ftpot}
\end{eqnarray}
where $n({\bf r})\equiv |\phi({\bf r})|^2 $ is the density and 
in Eq.  (\ref{ftpot})
we used a convolution of the respective variables 
to Fourier space and where tilde denotes  
Fourier
transformations: \cite{jb,YY,fisch,PS}
\begin{align}
\tilde U_{dd} ({\bf k})&= \frac{4\pi}{3}3  a_{dd} \biggr[ 
\frac{3k_z^2}{k^2}-1 \biggr], \\
\label{ftwf}
\tilde n({\bf k})&= \exp\left[-\frac{k_\rho^2 d_\rho^2}{4}\right]\tilde n_{1D}(k_z).
\end{align}
The $k_x,k_y$ integrals in  (\ref{ftpot}) can now be done and
\begin{eqnarray}
H_{dd}&=&\frac{4\pi N}{3}\frac{3 a_{dd}}{2}\frac{1}{2\pi}\int_{-\infty}^{\infty} 
dk_z\tilde n_{1D}(k_z)
\tilde n_{1D}(-k_z)
\frac{1}{(2\pi)^2}\nonumber \\
&\times&\int_{-\infty}^{\infty}\int_{-\infty}^{\infty}dk_x dk_y
\left[\frac{3k_z^2}{k_\rho^2+k_z^2}  -1 \right]\exp\left[-\frac{k_\rho^2
d_\rho^2}{2}\right],\nonumber \\
&\equiv & \frac{N}{2}\frac{1}{2\pi}\int_{-\infty}^{\infty} dk_z\tilde n_{1D}(k_z) \tilde n_{1D}(-k_z)
V_{1D}(k_z),
\end{eqnarray}
where the 1D potential in Fourier space is 
\begin{align}
V_{1D}(k_z)= &\,
{2 a_{dd}}\int_{0}^{\infty}dk_\rho   k_\rho
\left[\frac{3k_z^2}{k_\rho^2+k_z^2}  -1 \right]\exp\left[-\frac{k_\rho^2
d_\rho^2}{2}\right], \nonumber \\
\equiv &\, \frac{2a_{dd}}{d_\rho^2} s_{1D}(\frac{k_z d_\rho}{\sqrt 2}), 
\end{align}
\begin{align}
s_{1D}(\zeta) = \int_0^\infty  d u\left[  \frac{3\zeta^2}{u+\zeta^2}-1\right]e^{-u}. \label{zeta}
%&=& -\frac{4\pi}{3}\frac{3 a_{dd}}{2\pi d_\rho^2}-3 a_{dd}\exp(\frac{d_\rho^2 k_z^2}
%{2})k_z^2 {\mbox{EIE}}(-\frac{1}{2}d_\rho^2k_z^2),\nonumber
\end{align}
%with EIE=ExpIntegralEi.

The 1D potential in
configuration space is 
\begin{align}  
& U_{dd}^{1D}(Z) = \frac{1}{2\pi}\int_{-\infty}^{\infty} dk_z e^{ik_z z} V_{1D}(k_z) \notag \\
& = \frac{6 a_{dd}}{(\sqrt 2
d_\rho)^3} \left[\frac{4}{3}\delta(\sqrt t) +2\sqrt t-\sqrt
\pi (1+2 t) e^t {\mbox{erfc}}(\sqrt t)\right], \label{1dpotx}  
\end{align}
where $t=[Z/(\sqrt 2 d_\rho)]^2, Z=\vert z-z'\vert$.
Similar, but not identical, 1D reduced potential was derived in \cite{SS,deu}, where the
$\delta$-function term was absent. 
To derive the effective
1D equation for the cigar-shaped {dipolar BEC}, we substitute the
ansatz (\ref{anz1d}) in Eq.   (\ref{gp3d}), multiply by the ground-state wave
function $\phi(\rho)$ and integrate in $\rho$ to get
the 1D equation 
\begin{align}\label{vgp1d}
& i\frac{\partial \phi_{1D}(z,t)}{\partial
t}=\biggr[-\frac{\partial_z^2}{2}+\frac{\Omega_z^2 z^2}{2}+ 
\frac{2 aN}
{
d_\rho^2}\vert\phi_{1D}\vert^2
\nonumber \\ &+ \frac{2a_{dd}N}{d_\rho^2}
\int_{-\infty}^{\infty}\frac{dk_z}{2\pi}e^{ik_z z}\tilde
n(k_z)s_{1D}(\frac{k_z d_\rho}{\sqrt 2})\biggr]\phi_{1D}(z,t) , \\
&\equiv  \biggr[-\frac{\partial_z^2}{2}+\frac{\Omega_z^2 z^2}{2}+
 \frac{2 a N\vert\phi_{1D}\vert^2}
{
d_\rho^2} \nonumber \\
& + N\int_{-\infty}^{\infty} U_{dd}^{1D}( Z )\vert\phi_{1D}({z'},t)\vert^2d{z'}
\biggr] \phi_{1D}(z,t).
%\Omega_\rho^2d_\rho^4=1&+&gn/(2\pi)+d_\sigma^2\int_{-\infty}^{\infty}\frac{dk_z}{2\pi}
%e^{ik_z z } \tilde n (k_z)V(k_z) 
\end{align}

\end{document}